\documentclass[11pt]{article}

\usepackage{titlesec}
\usepackage{url}

\titlespacing\section{0pt}{10pt plus 4pt minus 2pt}{2pt plus 2pt minus 2pt}
\titlespacing\subsection{0pt}{10pt plus 4pt minus 2pt}{2pt plus 2pt minus 2pt}

\input epsf

\begin{document}

\noindent {\LARGE\bf NASA \textsl{Kepler} Mission White Paper} \vspace{1.5mm}

\noindent {\Large\bf Asteroseismology of Solar-Like Oscillators in a
  2-Wheel Mission} \vspace{1mm}

\noindent {\small{W.~J~Chaplin$^{1,2}$, H.~Kjeldsen$^2$,
    J.~Christensen-Dalsgaard$^{2}$, R.~L.~Gilliland$^{3}$,
    S.~D.~Kawaler$^{4}$, S.~Basu$^{5}$, J.~De~Ridder$^{6}$,
    D.~Huber$^{7}$, T.~Arentoft$^2$, J.~Schou$^{8}$,
    R.~A.~Garc\'ia$^{9}$, T.~S.~Metcalfe$^{10,2}$, K.~Brogaard$^{2}$,
    T.~L.~Campante$^{1,2}$, Y.~Elsworth$^{1,2}$, A.~Miglio$^{1,2}$,
    T.~Appourchaux$^{11}$, T.~R.~Bedding$^{12}$, S.~Hekker$^{8}$,
    G.~Houdek$^{2}$, C.~Karoff$^{2}$, J.~Molenda-\.Zakowicz$^{13}$,
    M.~J.~P.~F.~G.~Monteiro$^{14}$, V.~Silva~Aguirre$^{2}$,
    D.~Stello$^{12}$, W.~Ball$^{15}$, P.~G.~Beck$^{6}$,
    A.~C.~Birch$^{8}$, D.~L.~Buzasi$^{16}$, L.~Casagrande$^{17}$,
    T.~Cellier$^{9}$, E.~Corsaro$^{6}$, O.~L.~Creevey$^{11}$,
    G.~R.~Davies$^{1,2}$, S.~Deheuvels$^{18}$, G.~Do\u{g}an$^{19,2}$,
    L.~Gizon$^{8,15}$, F.~Grundahl$^{2}$, J.~Guzik$^{20}$,
    R.~Handberg$^{1,2}$, A.~Jim\'enez$^{21}$, T.~Kallinger$^{22}$,
    M.~N.~Lund$^{2}$, M.~Lundkvist$^{2}$, S.~Mathis$^{9}$,
    S.~Mathur$^{10}$, A.~Mazumdar$^{23}$, B.~Mosser$^{24}$,
    C.~Neiner$^{24}$, M.~B.~Nielsen$^{15}$, P.~L.~Pall\'e$^{21}$,
    M.~H.~Pinsonneault$^{25}$, D.~Salabert$^{9}$,
    A.~M.~Serenelli$^{26}$, H.~Shunker$^{8}$, T.~R.~White$^{15,12}$}}

\vspace{1mm}

\noindent {\scriptsize (1)~Univ. of Birmingham, UK (2)~Stellar
  Astrophysics Centre, Aarhus Univ., Denmark (3)~Pennsylvania State
  Univ., USA (4)~Iowa State Univ., USA (5)~Yale Univ., USA
  (6)~K.U. Leuven, Belgium (7)~NASA Ames Research Center, USA (8)~Max
  Planck Institute for Solar System Research, Germany
  (9)~CEA/DSM-CNRS-Univ. Paris Diderot, France (10)~Space Science
  Institute, USA (11)~Univ. Paris-Sud, Institut d'Astrophysique
  Spatiale, France (12)~Univ. of Sydney, Australia (13)~University of
  Wroc\l{}aw, Poland (14)~Univ. do Porto, Portugal (15)~Univ. of
  Goettingen, Germany (16)~Florida Gulf Coast Univ., USA (17)~The
  Australian National Univ., Australia (18)~Univ. de Toulouse, France
  (19)~HAO, Boulder, USA (20)~Los Alamos National Laboratory, USA
  (21)~IAC, Tenerife, Spain (22)~Univ. of Vienna, Austria (23)~Homi
  Bhabha Centre for Science Education, India (24)~Observatoire de
  Paris, France (25)~Ohio State Univ., USA (26)~Institute of Space
  Sciences (IEEC-CSIC), Spain}

\vspace{2.5mm}

{\centerline{{\bf Abstract}}}

This document is a response to the \textsl{Kepler} Project
\textsl{Call for White Papers}. In it, we comment on the potential for
continuing asteroseismology of solar-type and red-giant stars in a
2-wheel \textsl{Kepler} Mission. These stars show rich spectra of
solar-like oscillations.  Our main conclusion is that by targeting
stars in the ecliptic it should be possible to perform high-quality
asteroseismology, as long as favorable scenarios for 2-wheel pointing
performance are met. Targeting the ecliptic would potentially
facilitate unique science that was not possible in the nominal
Mission, notably from the study of clusters that are significantly
brighter than those in the \textsl{Kepler} field.

Our conclusions are based on predictions of 2-wheel observations made
by a space photometry simulator, with information provided by the
\textsl{Kepler} Project used as input to describe the degraded
pointing scenarios.  We find that elevated levels of
frequency-dependent noise, consistent with the above scenarios, would
have a significant negative impact on our ability to continue
asteroseismic studies of solar-like oscillators in the \textsl{Kepler}
field. However, the situation may be much more optimistic for
observations in the ecliptic, provided that pointing resets of the
spacecraft during regular desaturations of the two functioning
reaction wheels are accurate at the $\le 1\,\rm arcsec$ level. This
would make it possible to apply a post-hoc analysis that would recover
most of the lost photometric precision.  Without this post-hoc
correction---and the accurate re-pointing it requires---the
performance would probably be as poor as in the \textsl{Kepler}-field
case. Critical to our conclusions for both fields is the assumed level
of pointing noise (in the short-term jitter and the longer-term
drift). We suggest that further tests will be needed to clarify our
results once more detail and data on the expected pointing performance
becomes available, and we offer our assistance in this work.

\section{Introduction}
\label{sec:intro}

\subsection{Results from \textsl{Kepler}}

\noindent Asteroseismology has been one of the major successes of
\textsl{Kepler}. The research has been conducted in the framework of
the \textsl{Kepler} Asteroseismic Science Consortium (KASC; Gilliland
et al. 2010). \textsl{Kepler} has provided data of exquisite quality
for the asteroseismic study of unprecedented numbers of low-mass
main-sequence stars and cool subgiants (Chaplin et al. 2011a) and red
giants (e.g., Bedding et al. 2011; Hekker et al., 2011; Huber et
al. 2011; Kallinger et al., 2012; Mosser et al. 2012a; Stello et
al. 2013), including red giants in open clusters (e.g., Basu et
al. 2010; Stello et al. 2010; Corsaro et al. 2012; Miglio et
al. 2012). These stars show rich spectra of overtones of radial and
non-radial solar-like oscillations, pulsations that are stochastically
excited and intrinsically damped by near-surface convection. The
\textsl{Kepler} data have provided a unique combination of extremely
high-quality photometry and continuous, long-term coverage lasting up
to several years for many stars.  Long, high-duty-cycle lightcurves
provide the frequency resolution needed to measure the frequencies,
frequency splittings (from rotation and magnetic fields), and other
parameters of individual modes. These parameters allow estimation of
fundamental stellar properties (e.g., see Metcalfe et al. 2010, 2012),
the internal rotation (e.g., Mosser et al. 2012b)---including as a
function of radius (Beck et al. 2012; Deheuvels et al. 2012)---and are
unique probes of internal hydrostatic structure and stellar interiors
physics. Other examples include diagnostics of internal mixing and
small stellar cores (Silva~Aguirre et al. 2013), and ``acoustic
glitch'' signatures (Mazumdar et al. 2013) which allow estimation of
depths of convective envelopes (crucial information for dynamo
modellers) and potentially also stellar envelope helium abundances in
cool stars.

Even when the signal-to-noise ratios are of insufficient quality to
yield good values of individual frequencies, it is still possible to
extract estimates of so-called global or average asteroseismic
parameters, e.g., the large frequency separation of the oscillations
spectrum, and the frequency of maximum oscillation power (e.g., see
Huber et al. 2013). These global parameters may then be used to
provide estimates of the fundamental stellar properties (the precision
in the properties being inferior to what is possible with individual
frequencies, most notably for age).

Long-term coverage also opens the possibility to detect seismic
signatures of stellar activity and stellar activity cycles (Karoff et
al. 2009; Garc\'ia et al. 2010), which may be used in combination with
signatures of activity and surface rotation measured directly from the
lightcurves (from rotational modulation of starspots and active
regions) to further our understanding of the operation of dynamo
action in cool stars.

In cases where oscillations are detected in planet-hosting stars,
asteroseismology may be used to characterize the host star and
therefore also the properties of the orbiting planets (e.g., see
Carter et al. 2012; Barclay et al. 2013; Gilliland et al. 2013; Huber
et al. 2013). From high-frequency-resolution, high-quality data it is
also possible to use parameters of the rotationally split non-radial
modes to constrain the orientation of the spin axis of the star. In
systems with planets discovered by the transit method, this serves to
provide constraints on spin-orbit alignment, for understanding the
dynamic evolution of the systems (Chaplin et al. 2013).  Long-term
asteroseismic data on exoplanet hosts may also provide inferences on
stellar activity and stellar cycles, which is relevant to
understanding the influence that host stars have on their local
environments (with the obvious implications for planet habitability).

\subsection{New observations}
\label{sec:new}

\noindent These science drivers and data requirements motivate the
need for long-term, continuous coverage of the stars. Were
continuation of asteroseismology to be possible in the \textsl{Kepler}
field-of-view, extension of the baseline on existing targets would be
a natural choice. This would be particularly important for resolving
the hard-to-measure rotational signatures in main-sequence stars,
disentangling the complex spectra of gravity dominated modes in red
giants, and resolving modes in long-period giants (at the tip of the
red-giant branch, and in the asymptotic red-giant branch).

A switch of fields, combined with the possibility to continue to
collect high-quality asteroseismic data, would open the possibility to
target fresh cohorts of stars (e.g., for asteroseismic
characterization of newly discovered exoplanet hosts, assuming
continuation of exoplanet searches).  New targets in open clusters or
binaries would provide strong constraints for modelling, and would
make full use of the diagnostic potential of asteroseismology to test
stellar interiors physics. Such targets would also provide independent
data for testing the accuracy of asteroseismic estimates of
fundamental stellar properties. It is worth noting that the four open
clusters in the \textsl{Kepler} field are quite faint, and while they
have provided high-quality asteroseismic data on red giants, the
solar-type stars are too faint to yield asteroseismic
detections. Potentially interesting targets in the ecliptic (see
Section~\ref{sec:eclip}) include M67 and M44, which are not only
significantly brighter than the current \textsl{Kepler}
clusters---potentially making cluster asteroseismology of
main-sequence stars possible---but they also fill important age gaps
in the existing set. It might also be possible to target very bright
stars, as has been done in the \textsl{Kepler} field (although this
would depend critically on the sizes of apertures needed to track
stars in a 2-wheel scheme). For example, the binary 16\,Cyg contains
two solar analogues, with exquisite asteroseismic data that are
matched only by data on the Sun (Metcalfe et al. 2012). Indeed, the
height-to-background ratios of modes in the oscillation spectra of
these stars are limited by intrinsic stellar noise (from signatures of
granulation), and not shot or instrumental noise. Finally, a switch of
fields would also potentially give new asteroseismic data on red
giants for stellar population studies (e.g., see Miglio 2012), adding
to the data already available in the \textsl{Kepler} field-of-view and
the fields observed by CoRoT. Data in the \textsl{Kepler} field are
currently being exploited for population studies as part of the
APOKASC collaboration between APOGEE (part of SDSS~III) and KASC, with
APOGEE providing spectroscopic follow-up on large numbers of targets
(Pinsonneault et al., in preparation); the next generation of SDSS
could potentially provide similar follow-up on new fields.

\subsection{Layout of this paper}

\noindent The rest of our White Paper is laid out as follows. We begin
in Section~\ref{sec:nominal} by showing some basic performance metrics
from the nominal, fine-point \textsl{Kepler} data that are relevant to
asteroseismic signal-to-noise and quality. Our aim is to give a
baseline against which we can compare potential performance scenarios
in a 2-wheel configuration. In Section~\ref{sec:sim} we introduce
simulations of 2-wheel \textsl{Kepler} photometry.  Having results on
the frequency dependence of the noise is critical to judging the
potential to continue asteroseismic studies of solar-like oscillators.
Our preliminary conclusions from these simulations are discussed in
Section~\ref{sec:res}, where we comment on two scenarios: continuation
of observations in the \textsl{Kepler} field-of-view; and switching to
new fields in the ecliptic, a scenario that provides favorable
conditions for the 2-wheel configuration. Finally, we provide summary
remarks in Section~\ref{sec:conc}.

\section{Performance in the nominal Mission}
\label{sec:nominal}

\begin{figure}
 \centerline{\epsfxsize=8.5cm\epsfbox{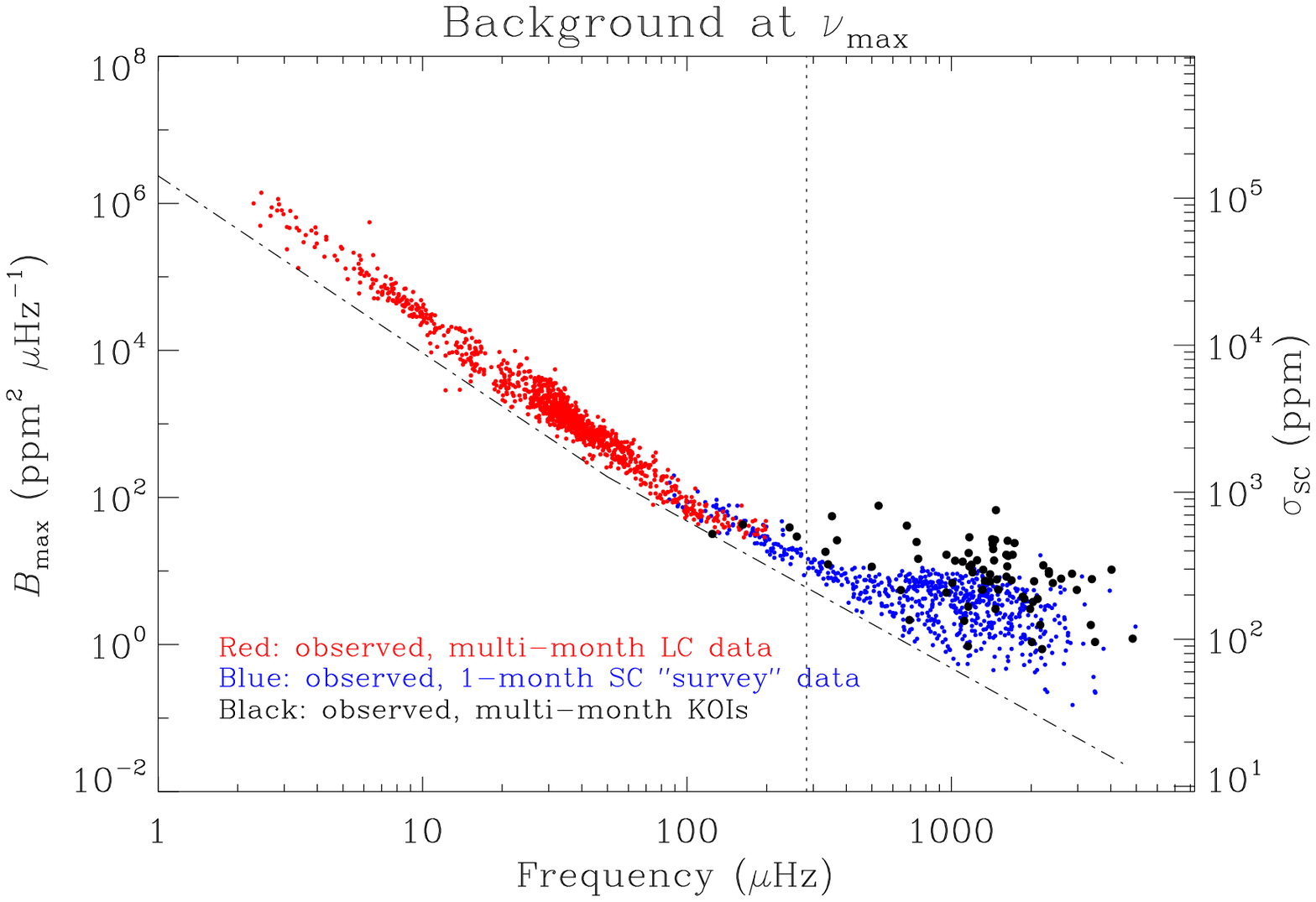}
             \epsfxsize=8.5cm\epsfbox{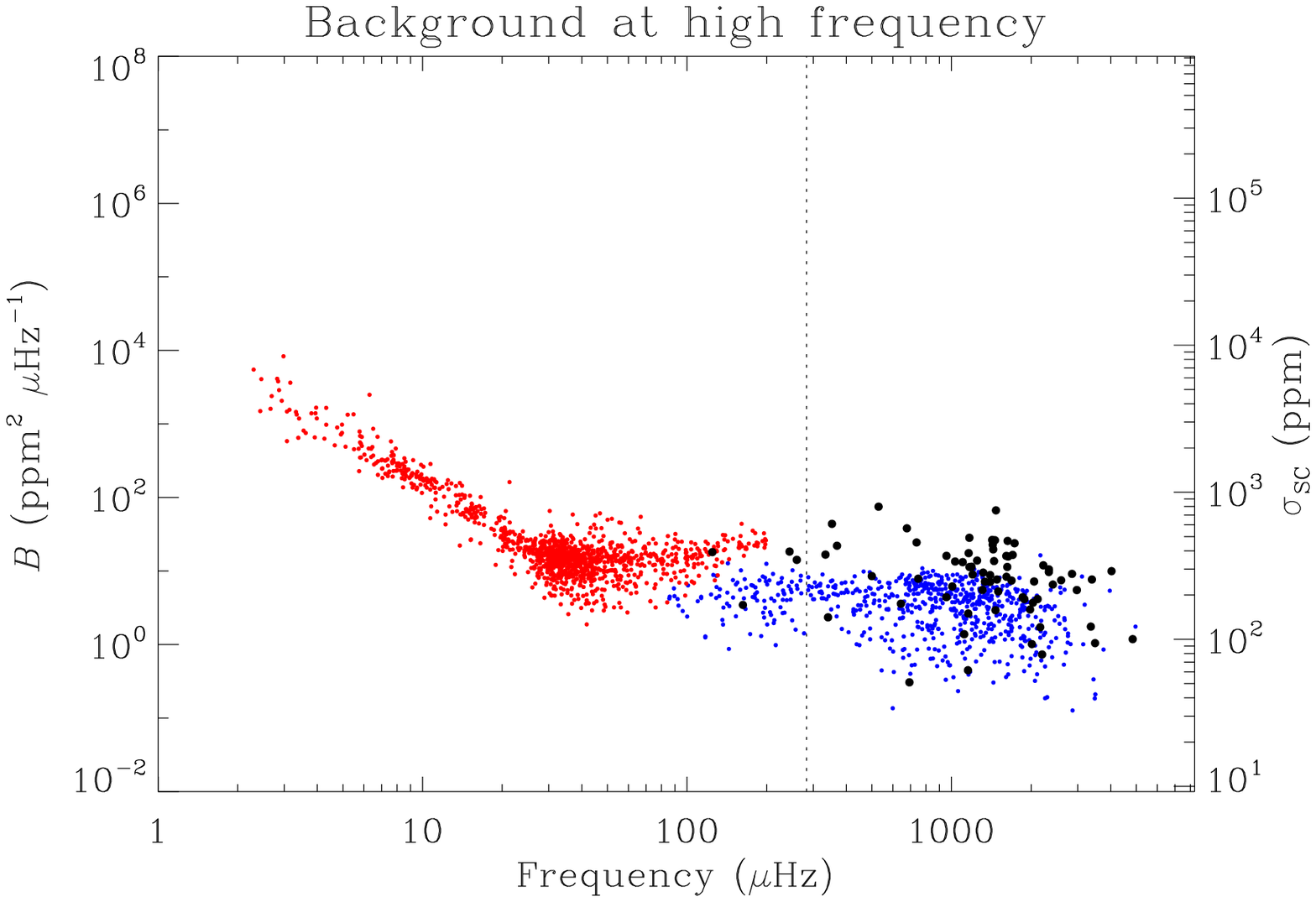}}
 \centerline{\epsfxsize=10.5cm\epsfbox{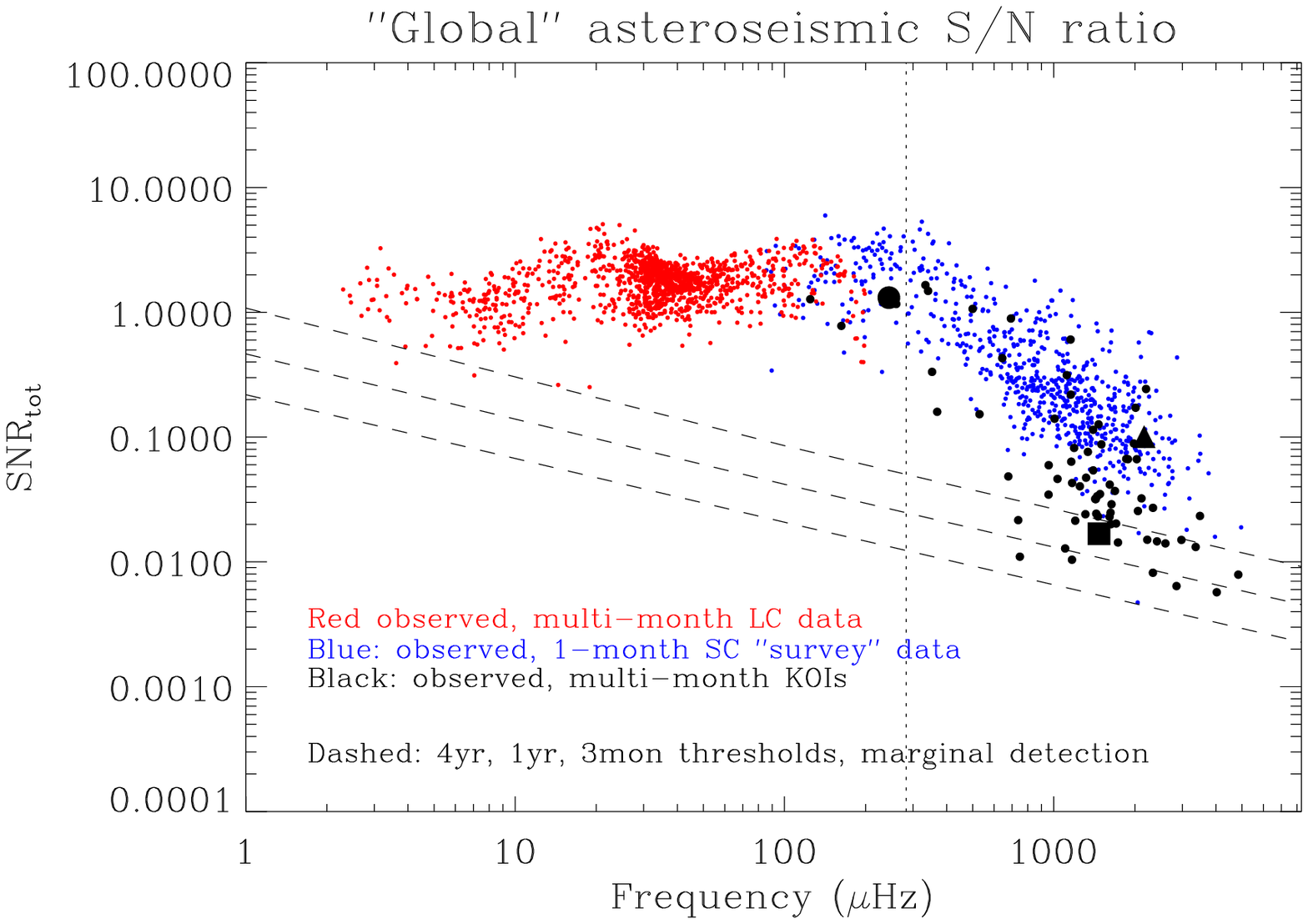}}
 \centerline{\epsfxsize=5.5cm\epsfbox{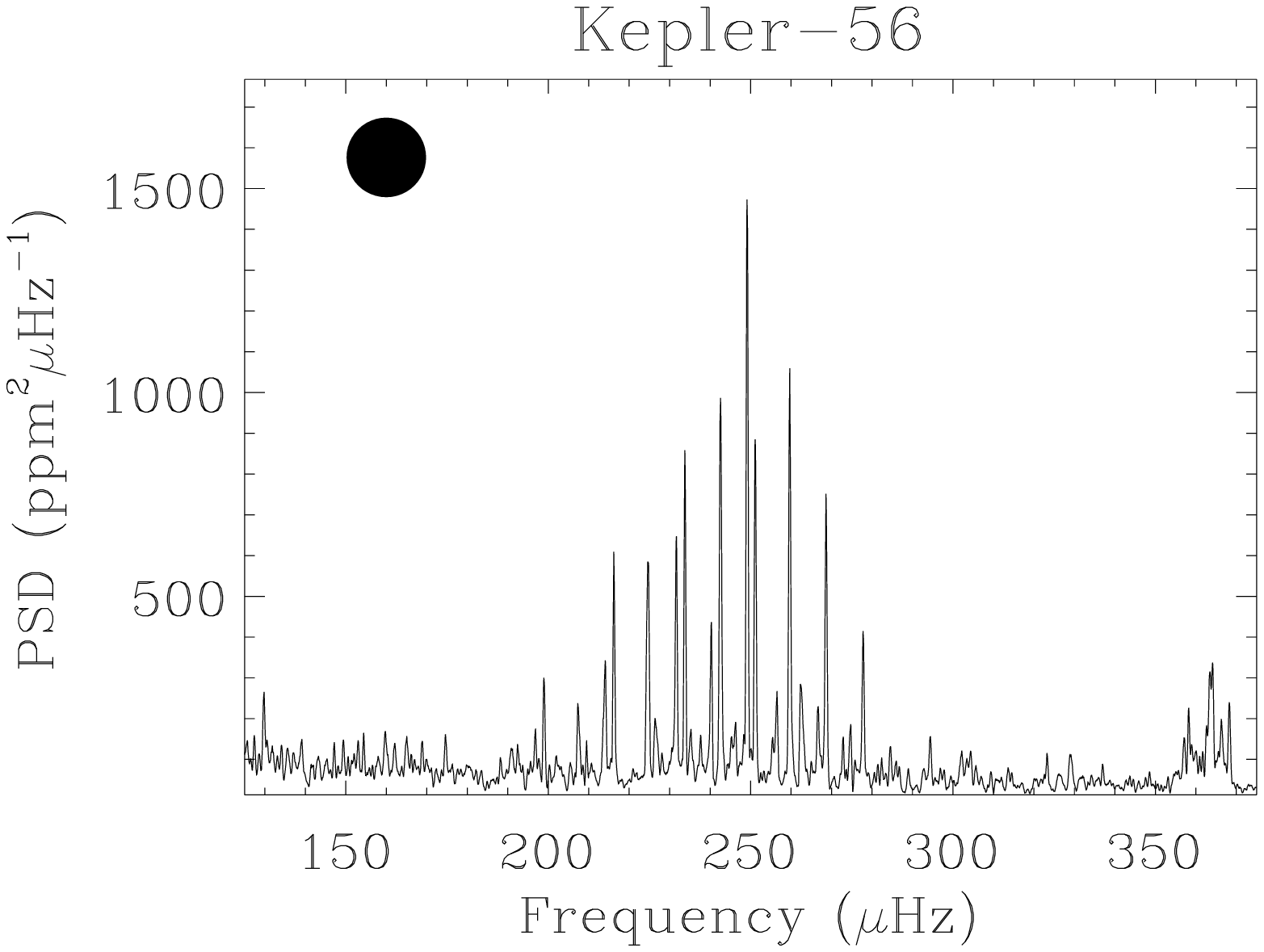}
             \epsfxsize=5.5cm\epsfbox{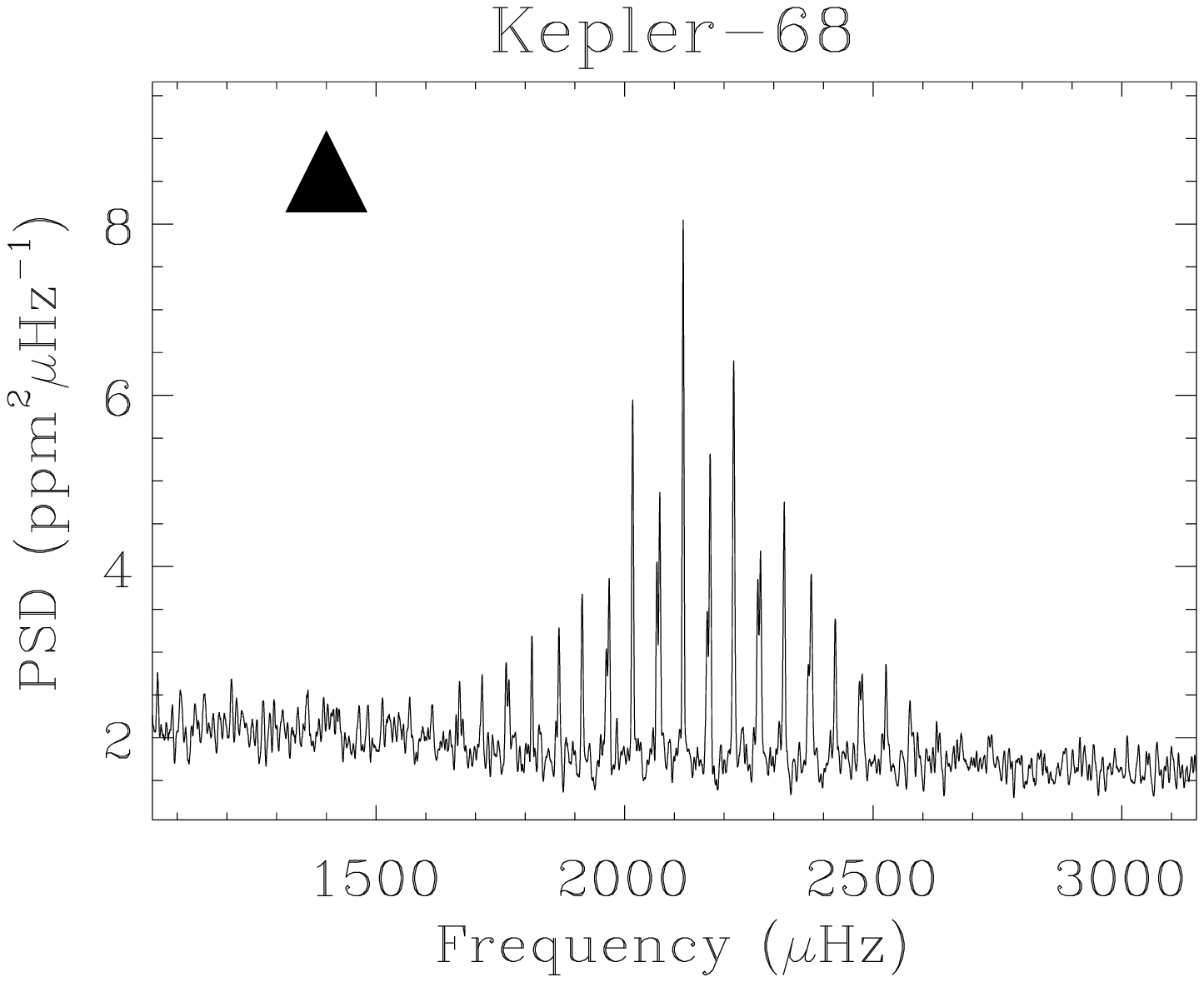}
             \epsfxsize=5.5cm\epsfbox{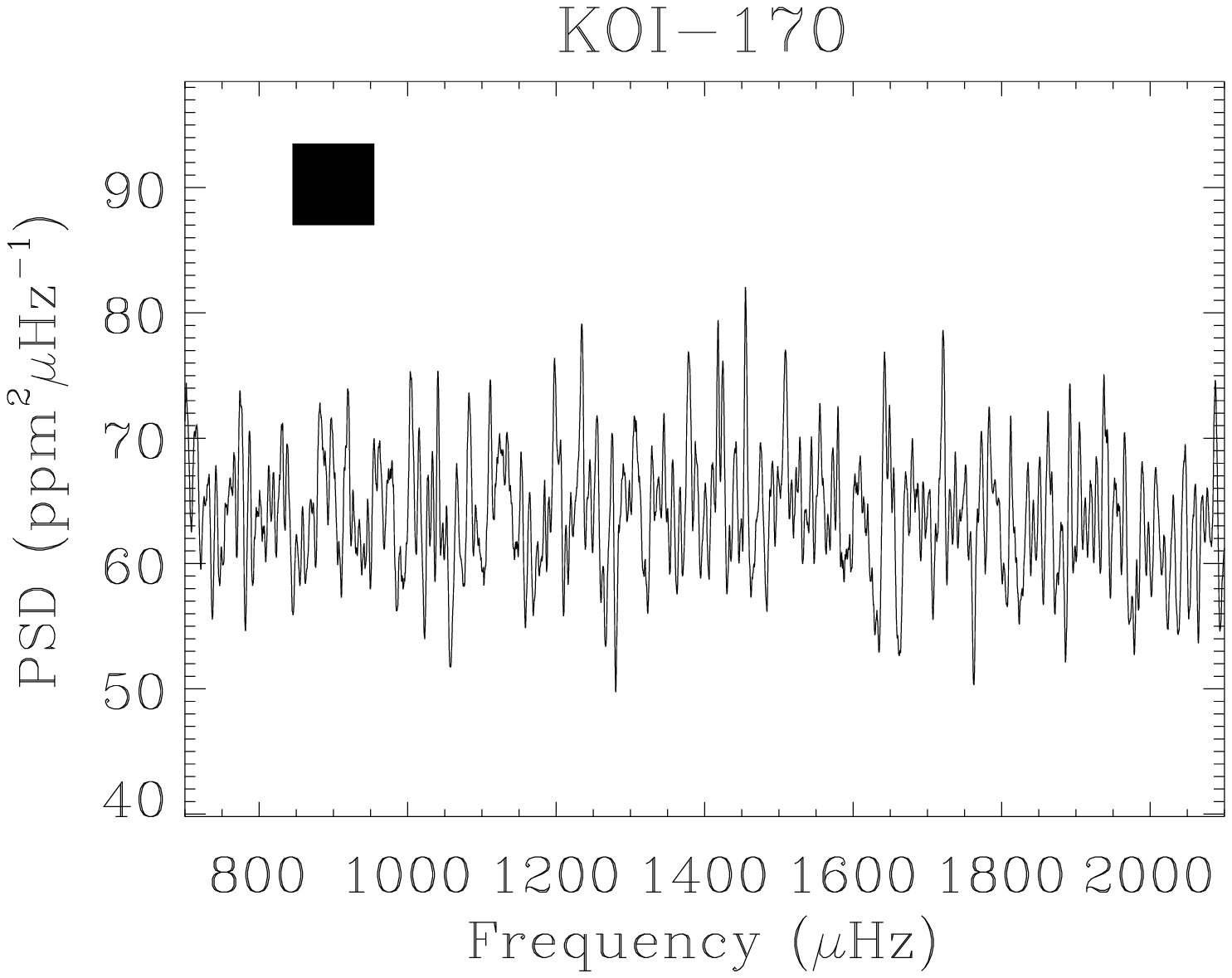}}
 
 \caption{\small Top left-hand panel: measured background power
   spectral density, $B_{\rm max}$, at the frequency of maximum
   oscillation power, $\nu_{\rm max}$, for three cohorts of solar-like
   oscillators observed by \textsl{Kepler} (see text and annotation
   for details). The right-hand axis shows the equivalent \textsc{rms}
   noise per 58.85-sec short-cadence (SC) sample. Top right-hand
   panel: measured backgrounds at high frequency, which are dominated
   by shot noise. Middle panel: asteroseismic global signal-to-noise
   ratios $\rm SNR_{tot}$ for stars in the target cohorts plotted in
   the top panel. The dashed lines show guideline ratios required for
   marginal detection of the oscillations assuming datasets of length
   3\,months (highest lying line), 1\,year, and 4\,years (lowest lying
   line).  Vertical dotted lines mark the Nyquist frequency for
   long-cadence observations. Bottom panels: Oscillation spectra of
   three KOIs, with their $\rm SNR_{tot}$ indicated by the matching
   symbols in the middle panel.}

 \label{fig:nominal}
\end{figure}

\noindent Fig.~\ref{fig:nominal} captures essential information on
shot, instrumental and intrinsic stellar noise, and the asteroseismic
signal-to-noise ratios achieved, in data from the nominal fine-point
\textsl{Kepler} Mission.


The top left-hand panel shows the measured background power spectral
density, $B_{\rm max}$, at the frequency $\nu_{\rm max}$ where the
oscillations present their maximum observed power.  The right-hand
axis shows the equivalent \textsc{rms} noise per 58.85-sec
short-cadence (SC) sample. We show results for all stars in three
cohorts of solar-like oscillators observed by \textsl{Kepler}: KASC
solar-type targets, KASC red-giant targets, and \textsl{Kepler}
Objects of Interest (predominantly solar-type stars, with a few red
giants). More details on the cohorts are given below. The plotted
$B_{\rm max}$ have contributions from shot and instrumental noise, as
well as intrinsic stellar noise. The right-hand panel shows the
measured backgrounds at high frequency. These backgrounds are
dominated by shot noise, and so reflect the native precision of the
\textsl{Kepler} photometry. In sum, both plots show background noise
levels that allowed detailed asteroseismic studies to be performed
using the nominal-Mission data.

The middle panel of Fig.~\ref{fig:nominal} shows ``global''
asteroseismic signal-to-noise ratios, $\rm SNR_{tot}$, for stars in
the target cohorts plotted in the top panel. We calculated these
ratios by dividing the total power observed in the oscillations
spectrum of each star by the integrated background power across the
frequency range occupied by the observed oscillations (see Chaplin et
al. 2011b)\footnote{Another definition commonly used is to divide the
  maximum power spectral density of the heavily smoothed envelope of
  oscillation power by $B_{\rm max}$, to give a
  ``height-to-background'' ratio at $\nu_{\rm max}$, e.g., Mosser et
  al. (2012c). Assuming a Gaussian-like power envelope, as is typical
  for many solar-like oscillators, this ratio is approximately a
  factor-of-two higher than $\rm SNR_{tot}$.}. It is important to note
that values of $\rm SNR_{tot}$ can be much less than unity and still
yield a prominent detection of the oscillations. The dashed lines show
guideline ratios required for marginal detection of the oscillations
assuming datasets of length 3\,months (highest lying line), 1\,year,
and 4\,years (lowest lying line). These thresholds are based on a
false-alarm test (again, see Chaplin et al. 2011b), which depends in
part on the number of frequency bins occupied by the oscillation
spectrum. The spectra are much narrower at low frequencies, and this
is what drives the higher thresholds needed there for detection (the
requirements are probably unduly pessimistic for the very lowest
frequencies). Finally, the bottom panels show the oscillation spectra
of three \textsl{Kepler} Objects of Interest (see below), with their
$\rm SNR_{tot}$ indicated by the matching symbols in the middle panel.

Details on the three cohorts are as follows. The first cohort
comprises more than 600 solar-type stars with detected oscillations,
and results on these stars are plotted in blue. These stars were
selected by the \textsl{Kepler} Asteroseismic Science Consortium
(KASC) to be observed for one month each during an asteroseismic
survey conducted during the first 10\,months of science
operations. The 58.85-sec SC data were needed to detect oscillations
in these cool main-sequence and subgiant stars, since the dominant
oscillations have periods of the order of minutes. These short periods
are not accessible to the 29.4-minute long-cadence (LC) data, for
which the Nyquist frequency is $\simeq 283\,\rm \mu Hz$ (marked here
by the vertical dotted line in both panels). These SC targets all have
\textsl{Kepler} apparent magnitudes brighter than $K\rm p \simeq 12$,
with one quarter brighter than $K\rm p \simeq 10$.  Around 150 of the
stars were subsequently observed in SC for at least one observing
quarter or longer, with around 60 targets having continuous multi-year
data from Q5 onwards. The second cohort comprises targets in black,
which are asteroseismic \textsl{Kepler} Objects of Interest (KOIs),
i.e., candidate, validated or confirmed exoplanet host stars showing
detected solar-like oscillations. All but a handful of the targets are
solar-type stars that were observed in SC. Many were put on long-term
SC observations because of their potential to yield asteroseismic data
for host-star characterization (based on asteroseismic detection
predictions made using the procedures described by Chaplin et
al. 2011b). Most of these datasets are therefore at least a year in
length. This target cohort also extends to fainter apparent magnitudes
than the KASC SC cohort, i.e., some stars were observed in the
expectation that they would yield only a marginal, but nevertheless
extremely useful, asteroseismic detection. The third cohort comprises
stars plotted in red, which are KASC red-giant targets. Red giants
pulsate at longer periods than solar-type stars, and their
oscillations are therefore accessible in the LC data. We note that in
addition to these KASC targets, solar-like oscillations have to date
also been detected in around another 14,000 \textsl{Kepler}
targets. The KASC targets have apparent magnitudes $K\rm p \simeq 11$
to 12; the other cohorts extend down to $K\rm p \simeq 14$.

In all but the faintest solar-type stars it is the contribution from
granulation which dominates the background $B_{\rm max}$.  The
dot-dashed line in the top left-hand panel of Fig.~\ref{fig:nominal}
is a simple model of the background contribution due to granulation
(see Chaplin et al. 2011b; Mosser et al. 2012c).  The spread in the
backgrounds for the solar-type stars reflects the significant
contribution from photon shot noise in many of these targets (in
particular for the faintest KOIs).  The middle panel of
Fig.~\ref{fig:nominal} also shows the low $\rm SNR_{tot}$ shown by
some of the KOIs (which required long datasets to yield an
asteroseismic detection).

Going to lower frequencies, the background contribution due to stellar
granulation increases in strength, and the distribution of plotted
points narrows in power. The $\rm SNR_{tot}$ increase toward lower
frequencies, through the subgiant phase. This is due to the
diminishing relative contribution to the background from shot noise,
whilst at the same time the power due to oscillations and granulation
increases. Indeed, the eventual flattening of the ratios can be
understood in terms of energy equipartition (the near-surface
convection excites the oscillations, so power in the granulation and
oscillations might be expected to change in the same way). The
increased density of points at around $\simeq 30\,\rm \mu Hz$ is due
to low-mass stars in the relatively long-lived He-core-burning
(red-clump) phase.

\section{Photometric simulations}
\label{sec:sim}

\noindent Our simulations of 2-wheel \textsl{Kepler} photometry were
made using the model developed by De~Ridder, Kjeldsen \& Arentoft
(2006). It simulates space-based photometric lightcurves given by CCD
images, with all significant instrumental effects included. Full
details of the simulations are given by Kjeldsen et al. (2013a, b)
[note that tested pointing scenarios have been updated since these
  documents were published]. Here, we summarize the main points.

Drift is the main source of instrumental noise at low frequencies, and
this is corrected almost entirely by the reaction wheels.  The
high-frequency attitude noise is a result of spacecraft motion that
cannot be removed by use of the guiding sensors or corrected by use of
the reaction wheels. In the simulations we therefore kept the
high-frequency noise component unchanged (as per the specifications in
Fig.~18 of the \textsl{Kepler Instrument Handbook}), but increased the
low-frequency noise to simulate the degraded performance of the
\textsl{Kepler} attitude in 2-wheel mode.

An important part of the simulation is the CCD sensitivity
variation. There are three major components that are used to describe
the relevant sensitivity variations:
 \begin{itemize}
 \itemsep -0.3em 
 \item[--] Global pixel-to-pixel (inter-pixel) variations: In the
   simulations we assumed a 5\,\% peak-to-peak sensitivity
   variability. This variability was chosen to be consistent with the
   1\,\% standard deviation in response non-uniformity given in the
   \textsl{Kepler Instrument Handbook}. We assumed that 80\,\% of the
   global pixel-to-pixel variation can be removed via flat fielding,
   as discussed in the Section~4.14 of the \textsl{Kepler} Instrument
   Handbook.
 \item[--] Random intra-pixel variations: We adopted a 25\,\% peak-to-peak
   variability within each pixel (on a scale corresponding to 1\,\% of
   the pixel area).
 \item[--] Sensitivity drops due to pixel-channel (i.e., intra-pixel
   drops due to the gate structure of the CCD chip): We used 10\,\%
   drop along 10\,\% of a pixel, in both orthogonal directions on the
   CCD.
\end{itemize}
Having specified drift and jitter due to the degraded attitude
control---details of which are given in Section~\ref{sec:res}---we
used a typical \textsl{Kepler} point-spread-function to create
artificial CCD images, which could then be analyzed. The photometric
analysis was performed using three different apertures (defined in the
software). We also measured the position of the point-spread-function
on the simulated CCD, which could then be compared to the attitude
inputs. The simulated lightcurves we produced contained only the
predicted photometric variability due to the pointing noise, which
could then be compared to the performance metrics for the
nominal-Mission data presented in Fig.~\ref{fig:nominal}.

\section{Simulation results and comments}
\label{sec:res}

\noindent In the following we use the spacecraft coordinate system
described in the \textsl{Call for White Papers} document. The two
functioning reaction wheels will be used to control the Y and Z axes,
while leaving the X axis (which points in the direction of the
boresight) to drift.

\subsection{Observations in the ecliptic}
\label{sec:eclip}

\noindent We began by simulating observations which approximate an
extremely favorable observing configuration, that of observing
targets in the ecliptic with the boresight pointing approximately in
the direction (or anti-direction) of the velocity vector of
\textsl{Kepler} in its orbit. This configuration has the Sun in the
``balancing'' X-Y plane, which minimizes torques due to solar
radiation pressure (which would induce roll about the X axis).

We assumed a non-linear (accelerating) roll totalling 120\,arcsec per
day (to match approximately the predictions in the document
\textsl{Explanatory Appendix to the Call for White Papers}), with
short-term \textsc{rms} jitter of 1\,arcsec. This gives a few pixels
of drift for targets lying on the edge of the field of view (the case
we tested), less for pixels lying close to the centre. Momentum
management demands regular desaturation of the two functioning wheels,
and here we assumed a daily cycle, whereby the pointing of the
spacecraft was reset every day. Moreover, we also considered the most
favorable reset scenario, in which targets were assumed to go back
each day to follow identical pixel trails on the CCD array. Here, at
each simulated daily reset we simply moved targets back to the same
pixels on our simulated array (but with the jitter imposed
independently on each repeat).  It should be borne in mind that it may
be difficult in practice to achieve this level of re-pointing. Offsets
at the first pointing would also propagate to give elevated levels of
drift, over and above those assumed above. However, as we shall see
below, the raw results from this assumed scenario are sufficient to
allow us to reach conclusions on asteroseismic detectability levels in
less favorable cases; and, very importantly, they also allow us to
comment on prospects for continuation of observations in the
\textsl{Kepler} field-of-view (of which more below in
Section~\ref{sec:kepler}).

Finally---and of crucial importance for our main conclusions---we also
applied a post-hoc correction to the photometry by using the repeated
daily tracking across the same pixels to produce a model of the local
inter- and intra-pixel variability. Details of this procedure may be
found in Kjeldsen et al. (2013a). Assuming that the detector
sensitivity is stable in time, when a star moves back to the same
position on the CCD several times one may use the measured photometry
and the measured stellar positions on the CCD to generate a model
which separates the intra- and inter-pixel sensitivity from other
variations in time, to in principle leave the intrinsic noise and the
stellar variability. The quality of the model depends on several
factors, e.g., how often the star is moved back to the same position,
and the drift rate during exposures. SC data will offer better
possibilities for this correction than LC data.  It is important to
stress that our results from applying this procedure rest on the
assumed best-case re-pointing. They should therefore be regarded as an
approximate guide to the ``limiting'' ideal case for the
photometry. We have also applied the test to simulated SC, not LC,
data.

\begin{figure}[!t]
 \centerline{\epsfxsize=8.5cm\epsfbox{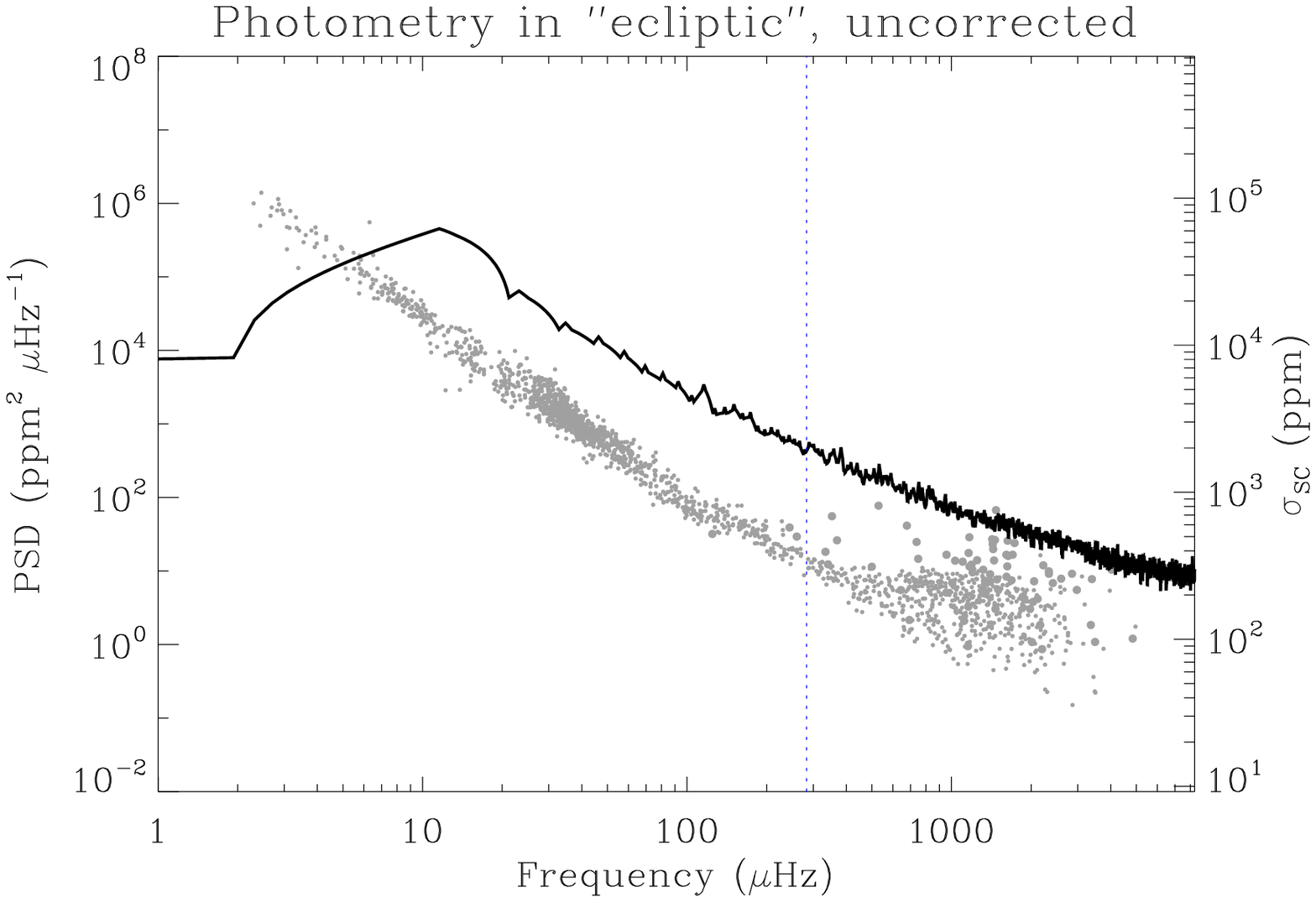}
             \epsfxsize=8.5cm\epsfbox{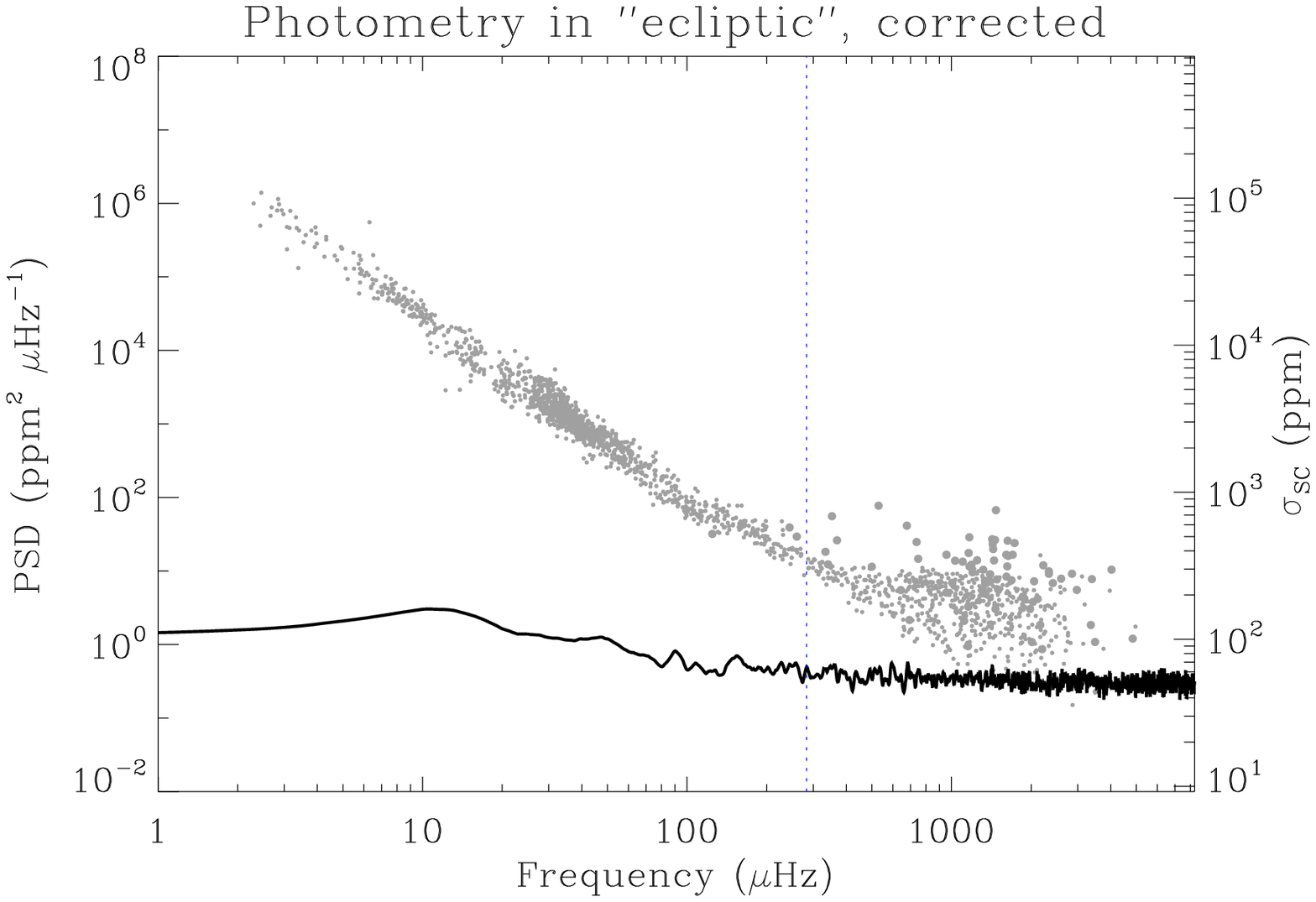}}
 \centerline{\epsfxsize=8.5cm\epsfbox{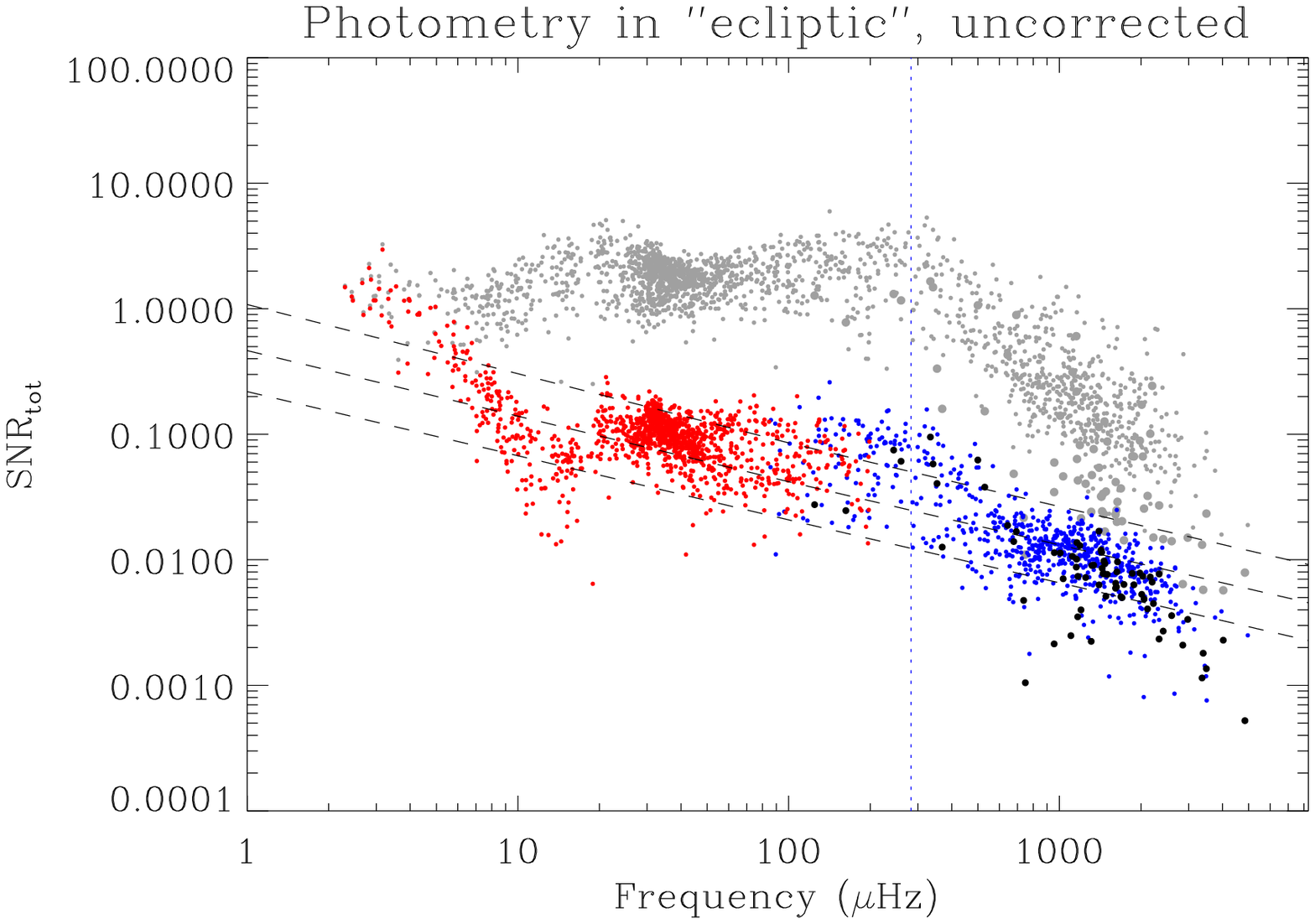}
             \epsfxsize=8.5cm\epsfbox{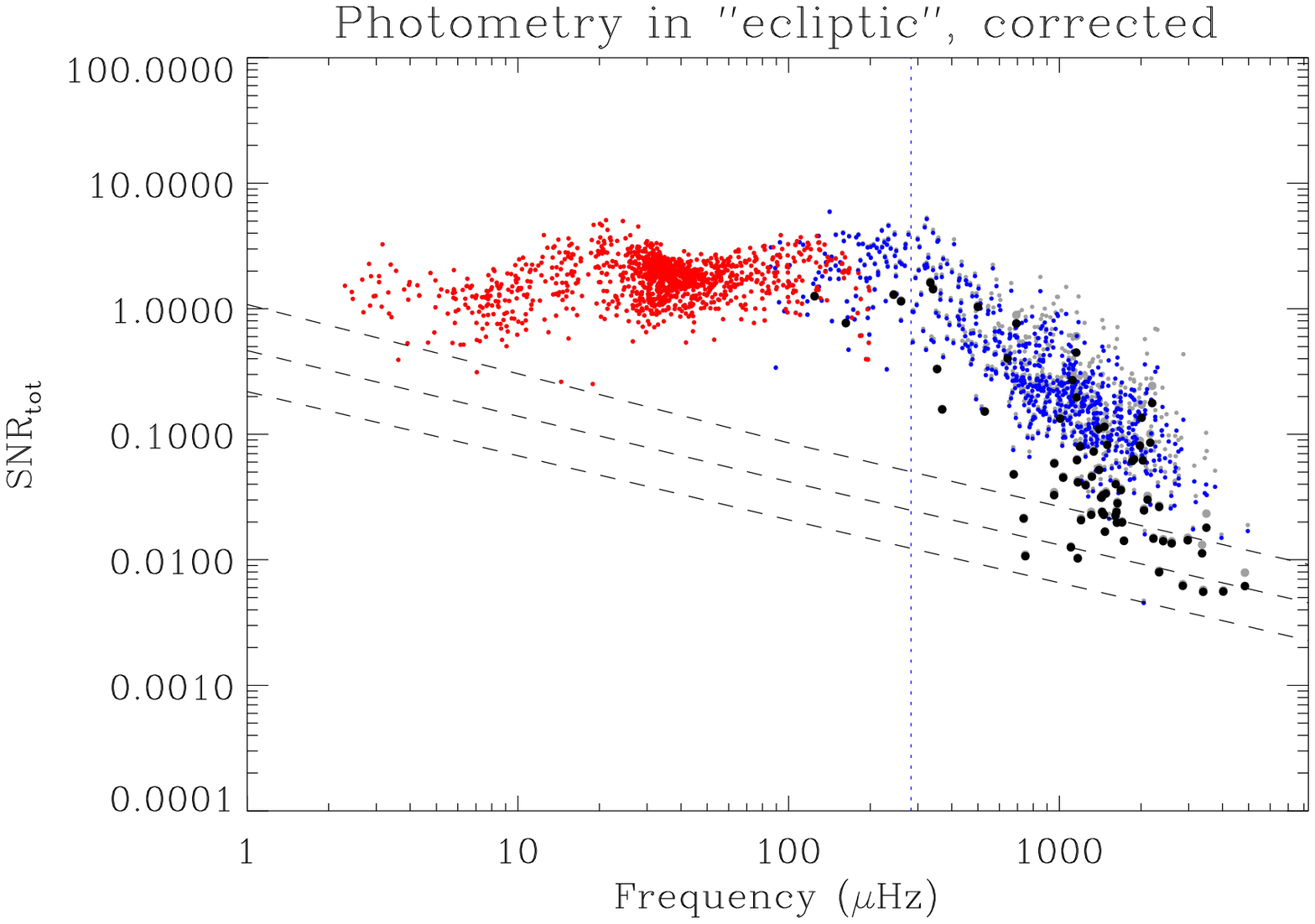}}

 \caption{\small Top left-panel: Power spectral density given by a
   simulated 30-day lightcurve of raw pointing noise. Top right-hand
   panel: Power spectral density after applying a post-hoc model to
   the raw data to correct for variations in CCD sensitivity. Both
   panels show in gray the $B_{\rm max}$ values for the
   nominal-Mission data (see top left-hand panel of
   Fig.~\ref{fig:nominal}). The right-hand axes show the equivalent
   \textsc{rms} noise per SC sample. Bottom left-hand panel:
   re-calculated $\rm SNR_{tot}$ with the effects of the raw pointing
   noise included. Gray points show ratios from nominal-Mission
   data. Right-hand panel: re-calculated ratios after post-hoc CCD
   corrections have been applied.}

 \label{fig:ecliptic}

\end{figure}

Fig.~\ref{fig:ecliptic} shows results from the simulations. The thick
black line in the top left-hand panel shows the power spectral density
given by a simulated 30-day lightcurve of raw SC pointing noise. The
spectrum has been smoothed for clarity with a 10-$\rm \mu Hz$ boxcar
filter. The thick line in the right-hand panel shows the smoothed
pointing-noise spectrum after applying a post-hoc model to the raw
data to correct for variations in CCD sensitivity. We also plot for
comparison (in gray) the $B_{\rm max}$ values for the nominal-Mission
data, which were plotted in the top left-hand panel of
Fig.~\ref{fig:nominal}. The scales on the right-hand axes again show
the equivalent \textsc{rms} noise per SC sample.

We may estimate the impact on the asteroseismic signal-to-noise ratios
$\rm SNR_{tot}$ by adding\footnote{Adding in power means this is an
  approximation, since we do not allow for interference with the noise
  (most important at marginal ratios).} the frequency-dependent
pointing noise power to the nominal-Mission $B_{\rm max}$. The bottom
left-hand panel shows the re-calculated ratios with the effects of the
raw pointing noise included; whilst the right-hand panel shows the
ratios after the post-hoc CCD corrections have been applied. Corrected
ratios are plotted in color; while the nominal-Mission ratios from
Fig.~\ref{fig:nominal} are shown in gray for comparison.


The \textsc{rms} scatter in the time domain of the raw pointing-noise
is about $2500\,\rm ppm$ per SC sample. The pointing noise has a
strong frequency dependence, and is higher than the nominal-Mission
backgrounds at essentially all frequencies. Not surprisingly, there
is also significant power at harmonics of 1\,day (suppressed visually
here by the smoothing applied for the plot). It is worth noting that
since the simulated lightcurve was only 30\,days long---and is but one
realization for a given track across the simulated CCD---there would
probably be significant variability in the pointing-noise power at very
low frequencies. Predictions there should be treated with caution.

The strongly degraded $\rm SNR_{tot}$ ratios in the bottom left-hand
panel show that the asteroseismic quality would be reduced
significantly, leading to null detections for many targets and perhaps
only marginal detections for others. Further tests indicated that the
simulated levels of noise---and hence the predicted outcomes for
asteroseismology---appear to depend critically on the assumed levels
of short-term jitter, in particular for the higher frequencies. The
results appear to be less sensitive to changes in the rate of drift
(assuming manageable rates).

The situation becomes rather more encouraging if one assumes that a
post-hoc model could successfully be applied to real 2-wheel
\textsl{Kepler} data (right-hand panels of Fig.~\ref{fig:ecliptic}).
The \textsc{rms} scatter of the corrected lightcurve---around $50\,\rm
ppm$ per SC sample---is significantly lower than that of the raw
lightcurve. The corrected pointing-noise background lies below the
nominal-Mission backgrounds for all but a few of the very brightest SC
targets (where the nominal-Mission precision was as low as a few-tens
of ppm). The corrected $\rm SNR_{tot}$ ratios in the bottom right-hand
panel are little changed from the nominal-Mission values, indicating
that asteroseismology of solar-like oscillators would in principle be
possible for targets in the ecliptic. It is important to remember that
this outcome would rest on the assumption that re-pointing is accurate
at the individual pixel level (i.e., probably demanding a $1\sigma$
precision of $\le 1\,\rm arcsec$).

Observations in the ecliptic would open exciting possibilities for new
targets, e.g., with reference to Section~\ref{sec:new}, the clusters
M67 and M44; and new ensembles of solar-type stars and red-giant field
stars not previously observed for asteroseismology, which would
provide data for stellar evolution and interiors physics studies, and
fundamental stellar properties for stellar population studies in new
fields, and exoplanet host-star characterization (assuming
continuation of exoplanet searches). Studies of the clusters could
enable science that was not possible in the nominal Mission (and which
would be unique): we would potentially be able for the first time to
exploit data on solar-like oscillations in main-sequence cluster
stars. Studies of surface rotation and activity, from variability in
the lightcurves, would probably also be viable on rapidly rotating
solar-type stars. Tests on longer timescales would be needed to judge
prospects for more slowly rotating solar-type stars and red giants.

Targets would not be observable all year round. One potential
observing strategy would have \textsl{Kepler} follow targets for
3\,months (with \textsl{Kepler} pointed exactly along the velocity
vector, say, half-way through the period); then re-point to another
field for 3\,months; and then go back to the original field (with
\textsl{Kepler} pointed against the velocity); and so on. This would
provide up to 6\,months of data on some stars every year.  Demands on
pointing would vary throughout each 3-month period.  It would be
highly desirable to have pixel tracks of the same length, which would
require a varying cadence for the desaturation events.  Again, our
results rest on the assumption of consistently achievable, very
accurate re-pointing. The most attractive strategy for
asteroseismology would then arguably be one in which \textsl{Kepler}
came back to the same targets every year, allowing the accumulation of
up to 1 or 2\,years of data after 2 or 4\,years of 2-wheel science
operations. Other choices would also of course be possible, depending
on the selection of specific targets in the ecliptic. Long, 3-month
gaps in coverage would not pose a significant obstacle to the analysis
of main-sequence stars. The lifetimes of modes observed in these stars
are typically much shorter, of the order of days. The lifetimes of
mixed modes and gravity dominated modes observed in more evolved stars
are somewhat longer; however, the effects of long gaps can in
principle be modelled and accounted for. There would also be repeated
desaturation gaps, but they should be short and not a significant
issue for the analysis.

\subsection{Observations in the \textsl{Kepler} field-of-view}
\label{sec:kepler}

\noindent Next, we comment on the extremely important scenario of
continuing observations in the \textsl{Kepler} field-of-view.  This
scenario has the added complication that the Sun drifts relative to
the X-Y plane of the spacecraft by 1\,degree per day. The spacecraft
must therefore be rolled on a regular basis to maintain the Sun in the
balancing X-Y plane. Frequent rolls (probably daily) would be required
to keep drift rates manageable, and to ensure they would not become
problematic for target aperture allocations. Targets will not return
to the same pixel tracks on the CCD array and it would therefore not
be possible to use the post-hoc procedure to model and then correct
the raw photometry.

To simulate the impact of an assumed daily roll, we simply reset
targets each day in our simulation to a different part of the
simulated array. It should also be noted that our simulations did not
include complications that would arise from changing stray-light
levels (from the daily rolls). These revised simulations gave results
that were similar to the uncorrected ecliptic case discussed above
(again indicating the importance of jitter, which was the same in both
sets of simulations, versus the impact of manageable drift, which was
not).  Bearing in mind the simulation results in the left-hand panels
of Fig.~\ref{fig:ecliptic}, the prospects for continuing
asteroseismology of solar-like oscillators in the \textsl{Kepler}
field do not look good (unless the pointing noise specifications we
have tested turn out to be unduly pessimistic).

\section{Concluding remarks}
\label{sec:conc}

\noindent We have performed our own simulations of the pointing noise
expected for 2-wheel \textsl{Kepler} observations of targets in the
\textsl{Kepler} field-of-view, and in the ecliptic. Having results on
the frequency dependence of the noise is critical to judging the
potential to continue asteroseismic studies of solar-like oscillators.
We find that while elevated levels of noise would impact significantly
on our ability to continue such studies in the \textsl{Kepler} field,
the situation is potentially much more optimistic for observations in
the ecliptic. However, that optimism rests on the important assumption
that pointing resets during regular desaturations of the two
functioning wheels would be accurate at the $\le 1\,\rm arcsec$
level. This would ensure that when targeting a given field, stars
would follow the same pixel tracks on the CCD after each desaturation
and reset, making it possible to apply a post-hoc analysis to recover
most of the photometric precision lost by the inferior pointing.  Our
simulations indicate that \textsl{Kepler} would then be able to
provide good data for asteroseismology.  Assuming good pointing
performance, it would probably be possible to use the existing readout
software with slightly increased aperture masks.

A crucial part of our paper is the set of ``baseline'' asteroseismic
performance metrics, which come from analysis results on the
nominal-Mission data. Once full simulation results on the actual
expected (frequency-dependent) performance are available---courtesy of
the \textsl{Kepler} Project and Ball Aerospace (the spacecraft
manufacturer)---it will be possible to use the metrics presented in
this paper (and accompanying data that we did not have the space to
include) to make a definitive assessment of whether continuation of
asteroseismic studies of solar-like oscillators is feasible. We will
also be able to test observations in other fields (including pointing
just off the ecliptic).

\vspace{3mm}
\noindent\textbf{References}
\vspace{1mm}

\scriptsize \noindent Barclay, T., et al., 2013, Nature, 494, 452

\noindent Basu, S., et al., 2010, ApJ, 729, L10

\noindent Beck, P. G., et al., 2012, Nature, 481, 55

\noindent Bedding, T. R., et al., 2011, Nature, 471, 608

\noindent Carter, J. A., et al., 2012, Science, 337, 556

\noindent Chaplin, W. J., et al., 2011a, Science, 332, 213

\noindent Chaplin, W. J., et al., 2011b, ApJ, 732, 54

\noindent Chaplin, W. J., et al., 2013, ApJ, 766, 101

\noindent Corsaro, E., et al., 2012, ApJ, 757, 190

\noindent Deheuvels, S., et al., 2012, ApJ, 756, 19

\noindent De~Ridder, J., et al., 2006, MNRAS, 365, 595

\noindent Garc\'ia, R. A., et al., 2010, Science, 329, 1032

\noindent Gilliland, R. L., et al., 2010, PASP, 122, 131

\noindent Gilliland, R. L., et al., 2013, ApJ, 766, 40

\noindent Hekker, S., et al., 2011, MNRAS, 414, 2594

\noindent Huber, D., et al., 2011, ApJ, 743, 143

\noindent Huber, D., et al., 2013, ApJ, 767, 127

\noindent Kallinger, T., et al., 2012, A\&A, 522, 1

\noindent Karoff, C., et al., 2009, MNRAS, 399, 914

\noindent Kjeldsen, H., et al., 2013a, KASC document, DASC/KASOC/0043(2)

\noindent (\url{http://astro.phys.au.dk/~hans/Call_for_White_Paper/DASC_KASOC_0043_2.pdf})

\noindent Kjeldsen, H., et al., 2013b, KASC document, DASC/KASOC/0044(1)

\noindent (\url{http://astro.phys.au.dk/~hans/Call_for_White_Paper/DASC_KASOC_0044_1.pdf})

\noindent Mazumdar, A., et al., 2013, ApJ, submitted

\noindent Metcalfe, T. S., et al., 2010, ApJ, 723, 1583

\noindent Metcalfe, T. S., et al., 2012, ApJ, 748, L10 

\noindent Miglio, A., 2012, ASSP (ISBN 978-3-642-18417-8), p.~11

\noindent Miglio, A., et al., 2012, MNRAS, 419, 2077

\noindent Mosser, B., et al., 2012a, A\&A, 540, 143

\noindent Mosser, B., et al., 2012b, A\&A, 548, 10

\noindent Mosser, B., et al., 2012c, A\&A, 537, 30

\noindent Silva~Aguirre, V., et al., 2013, ApJ, 769, 141

\noindent Stello, D., et al., 2010, ApJ, 713, L182

\noindent Stello, D., et al., 2013, ApJ, 765, L41

\noindent \textsl{Kepler Instrument Handbook}, KSCI-19033, 15 July 2009, NASA
Ames Research Center

\noindent (\url{http://keplerscience.arc.nasa.gov/calibration/KSCI-19033-001.pdf})

\noindent \textsl{Call for White Papers: Soliciting Community Input
  for Alternative Science Investigations for the Kepler Spacecraft}

\noindent (\url{http://keplergo.arc.nasa.gov/docs/Kepler-2wheels-call-1.pdf})

\noindent \textsl{Explanatory Appendix to the Kepler Project call for
  White Papers: Kepler 2-Wheel Pointing Control}

\noindent (\url{http://keplergo.arc.nasa.gov/docs/Kepler-2-Wheel-pointing-performance.pdf})

\end{document}